\documentclass[utf8]{frontiersSCNS} 

\usepackage{url,hyperref,lineno,microtype,subcaption}
\usepackage[onehalfspacing]{setspace}
\usepackage{aas_macros}

\def\mean#1{\left< #1 \right>}
\def\SF{S\negthinspace F}

\def\keyFont{\fontsize{8}{11}\helveticabold }
\def\firstAuthorLast{Serafinelli {et~al.}} 
\def\Authors{R. Serafinelli\,$^{1,2,*}$, F. Vagnetti\,$^{1}$, E. Chiaraluce\,$^{1}$, and R. Middei\,$^{3}$}
\def\Address{$^{1}$Dipartimento di Fisica, Universit\`a di Roma Tor Vergata, Roma, Italy \\
$^{2}$Dipartimento di Fisica, Universit\`a di Roma Sapienza, Roma, Italy \\
$^{3}$Dipartimento di Matematica e Fisica, Universit\`a Roma Tre, Roma, Italy  }
\def\corrAuthor{Roberto Serafinelli}

\def\corrEmail{roberto.serafinelli@roma2.infn.it}

\begin{document}
\onecolumn
\firstpage{1}

\title[Quasar X-ray Variability]{The MEXSAS2 Sample and the Ensemble X-ray Variability of Quasars} 

\author[\firstAuthorLast ]{\Authors} 
\address{} 
\correspondance{} 

\extraAuth{}

\maketitle

\begin{abstract}

\section{}
We present the second Multi-Epoch X-ray Serendipitous AGN Sample (MEXSAS2), extracted from the 6th release of the XMM Serendipitous Source Catalogue (XMMSSC-DR6), cross-matched with Sloan Digital Sky Survey quasar catalogues DR7Q and DR12Q. Our sample also includes the available measurements for masses, bolometric luminosities, and Eddington ratios. Analyses of the ensemble structure function and spectral variability are presented, together with their dependences on such parameters. We confirm a decrease of the structure function with the X-ray luminosity, and find a weak dependence on the black hole mass. We introduce a new spectral variability estimator, taking errors on both fluxes and spectral indices into account. We confirm an ensemble softer when brighter trend, with no dependence of such estimator on black hole mass, Eddington ratio, redshift, X-ray and bolometric luminosity.

\tiny
 \keyFont{ \section{Keywords:} Catalogs - Quasars - Spectral Variability - Structure Function - X-rays} 
 
\end{abstract}

\section{The MEXSAS2 Catalogue}
We present here the MEXSAS2 catalogue, obtained from the sixth release of the XMM Newton Serendipitous Source Catalogue \citep[XMMSSC-DR6, ][]{rose16}, combined with both SDSS-DR7Q \citep{schn10} and SDSS-DR12Q quasar catalogues \citep{pari17}. The MEXSAS2 catalogue is the updated version of MEXSAS (Multi-Epoch XMM Serendipitous AGN Sample), published by \citet{vagn16}. It contains 9735 X-ray observations for 3366 quasars, which increases by about 25\% the numbers of the previous version (7837 observations, 2700 quasars). It only contains sources with more than one observation, and it is therefore ideal for variability studies.\\
We complement the catalogue with measurements of black hole masses, bolometric luminosities and Eddington ratios, which are available for 3138 quasars (93\%) from the catalogues by \citet{shen11} for the sources listed in SDSS-DR7Q, and by \citet{kozo17} for SDSS-DR12Q. We use homogeneous criteria to derive the masses and bolometric luminosities from the two catalogues, adapting their criteria, which are slightly different for the redshift intervals where two different broad lines and two continuum luminosities are available. In fact, \citet{shen11} adopts different single-epoch virial estimates sharply dividing at redshift $z=0.7$ for choosing between H$\beta$ and MgII(2798\AA) relations, and at $z=1.9$ for Mg\,II(2798\AA) and C\,IV(1549\AA). A similar option is adopted for computing bolometric luminosities from the continua at 5100\AA\ or 3000\AA\, dividing at $z=0.7$, and at 3000\AA\ or 1350\AA\, dividing at $z=1.9$. The criteria adopted by \citet{kozo17} are instead to prefer the Mg\,II black hole mass estimate when available, rather than that from C\,IV, which is biased from various effects, and to compute bolometric luminosity as a weighted average of those derived from two different available continua. However, \citet{kozo17} also find that the DR7Q line widths (which are derived from detailed fits of spectral components) are generally more reliable than the DR12Q ones (which are derived from principal component analysis), thus they are to be preferred, when available. To apply the same criteria to the mass and luminosity data from both DR7Q and DR12Q, we adopt the following choice: (i) for quasars included only in DR12Q and not in DR7Q, we take the estimates by \citet{kozo17} (2178 objects); (ii) for quasars included in both catalogues or only in DR7Q, and in redshift intervals with only one broad line and one continuum luminosity available ($z\lesssim 0.35$, $0.9\lesssim z\lesssim 1.5$, $z\gtrsim 2.25$), we take the estimates by \citet{shen11} (442 objects); (iii) for quasars included in both catalogues or only in DR7Q, in redshift intervals with two broad lines and continua ($0.35\lesssim z\lesssim 0.9$, $1.5\lesssim z\lesssim 2.25$), we apply the \citet{kozo17} criteria to the DR7Q data, deriving new estimates (518 objects).\\
The catalogue, including X-ray measurements from XMMSSC and quasar data from SDSS and from our elaboration, will be published elsewhere (Vagnetti et al, in preparation). Here we provide preliminary results of our ensemble analyses of the X-ray variability.

\section{FLUX VARIABILITY AND STRUCTURE FUNCTION}
We compute the structure function according to \citet{vagn16}, as a r.m.s. difference of the X-ray flux measured at two epochs differing by $\tau$ in the rest-frame, and corrected for the noise contribution, 
\begin{equation}
\SF(\tau)\equiv\sqrt{\langle[\log f_X(t+\tau)-\log f_X(t)]^2\rangle-\sigma^2_{noise,SF}}\quad,
\end{equation}

\noindent
where $\sigma_{noise,SF}^2=\mean{\sigma_n^2(t)+\sigma_n^2(t+\tau)}$ is the quadratic contribution of the photometric noise to the observed variations, $\sigma_n$ being the error on $\log f_X$ at each given time.\\
We use mainly EPIC X-ray fluxes in the XMM-Newton band 9 (0.5-4.5 KeV), as in our previous papers \citep{vagn11,vagn16}.\\
The SF can be fitted by a power-law $\SF\propto\tau^b$. For the whole sample we find a slope $b=0.11\pm0.01$. The SFs in bins of X-ray luminosity and black hole mass are shown in Fig.~1. We confirm a strong anti-correlation with X-ray luminosity, approximately as $L_X^{-0.22}$, with correlation coefficient $r=-0.92$ and null probability $p(>r)=0.04$, and no dependence on redshift, similarly to \citet{vagn16}. There is an apparent decrease of the SF with black hole mass, approximately as $M_{BH}^{-0.15}$, with $r=-0.97$ and $p(>r)=0.03$, but partial correlation analysis suggests that this is due to the strong correlation of mass with X-ray luminosity. Limiting the analysis to the luminosity interval $10^{44}{\rm\,erg/s}<L_X<10^{45}{\rm\,erg/s}$, the mass-luminosity correlation reduces ($r=-0.83$, $p(>r)=0.06$), and the dependence of the SF on black hole mass is weaker, $\propto M_{BH}^{-0.06}$. We find neither dependence on bolometric luminosity nor on Eddington ratio. 

\section{SPECTRAL VARIABILITY}
We update the analysis of the spectral variability parameter, initially introduced by \citet{trev02} and recently adapted to the X-ray band by \citet{sera17}
\begin{equation}
\beta=-{\Delta\Gamma\over\Delta\log f_X}\quad,
\end{equation}
relating variations of the photon index $\Gamma$ with those of the X-ray flux in a given band.\\ 
The spectra of most Seyfert galaxies with Eddington ratios above 0.01 typically become steeper in their brighter phases \citep[e.g.][]{mark03,sobo09,conn16}, as well as for many galactic black hole binary systems \citep[e.g.][]{remi06,done07,dong14}. This is known as `softer when brighter' behaviour and translates to a negative $\beta$ value according to Eq. 2. We also found this `softer when brighter' behaviour, obtaining $\beta=-0.69\pm0.03$ in our ensemble analysis, for the previous version of the MEXSAS sample, using fluxes in the soft X-ray band 0.5-2 keV, and computing variations with respect to the mean values of each source \citep{sera17}.\\ 
Since the ensemble correlation between $\Delta\Gamma$ and $\Delta\log f_X$ contains a large scatter, also due to the large measurement errors for the fainter sources, we fit a linear relation $\Delta\Gamma\propto\Delta\log f_X$ to the MEXSAS2 data set taking the uncertainties in both variables into account. We run a high number of linear fits replacing original data with Gaussian distributed values within the associated error box. Moreover, following \citet{isob90}, we computed both ordinary least squares regressions OLS$(Y|X)$ of the dependent variable $Y$ on the independent variable $X$ (which result we call for brevity $\beta_{xy}$), the inverse regression OLS$(X|Y)$ ($\beta_{yx}$), and the bisector:
\begin{equation}
\beta_{bis}={\beta_{xy}\beta_{yx}-1+\sqrt{(1+\beta_{xy}^2)(1+\beta_{yx}^2)}\over \beta_{xy}+\beta_{yx}}\quad.
\end{equation}
For the whole sample, we find ensemble values $\beta_{xy}=-0.22\pm0.04$ and $\beta_{bis}=-1.22\pm0.05$, confirming the `softer when brighter' behaviour for both spectral variability parameters (see Fig.~2).\\
We then divide our sample in bins of soft X-ray flux, to show that the resulting $\beta$ does not change within the errors (see Fig.~3). In fact this was expected, because measurement errors do affect the estimate of $\beta$, but this must reflect an intrinsic relation between the flux and slope variations which should not change with an observational parameter like the average flux of the sources.\\
We also divide our sample in bins of black hole mass, $M_{BH}$, and show the result in Fig.~3. Both $\beta_{xy}$ and $\beta_{bis}$ are independent of $M_{BH}$ within the errors, and always compatible with their overall ensemble values.\\
We similarly analyse possible dependencies on Eddington ratio, redshift, X-ray and bolometric luminosities, finding no evidence of change. This analysis will be reported in an upcoming paper (Vagnetti et al., in prep.). 

\section{SUMMARY}
\begin{itemize}
\item[-] We have updated our X-ray quasar catalogue MEXSAS to MEXSAS2.
\item[-] We have confirmed our previous results for the structure function dependence on X-ray luminosity, and no dependence on redshift; we find a weak dependence on black hole mass, and no dependence on bolometric luminosity and Eddington ratio.
\item[-] We have developed new spectral variability estimators, by computing ordinary least squares and bisector fits with errors in both variables.
\item[-] We obtain `softer when brighter' trends for the new MEXSAS2 sample using both estimators.
\item[-] Our spectral variability estimates, that also take errors on both fluxes and spectral indices into account, do not present a dependence on the flux.
\item[-] No trend with black hole mass, Eddington ratio, redshift, X-ray and bolometric luminosities is found, in agreement with our previous results \citep{sera17}.
\end{itemize}

\section*{Author Contributions}
RS performed the spectral variability analysis and wrote part of the text, FV was responsible for the idea that resulted in the paper and also contributed to the text, EC built the dataset of the MEXSAS2 catalogue, RM performed the structure function analysis.

\bibliographystyle{frontiersinSCNS_ENG_HUMS} 
\bibliography{pd}


\section*{Figure captions}

\begin{figure}[h!]
\begin{minipage}[b]{.5\linewidth}
\centering\includegraphics[width=8cm]{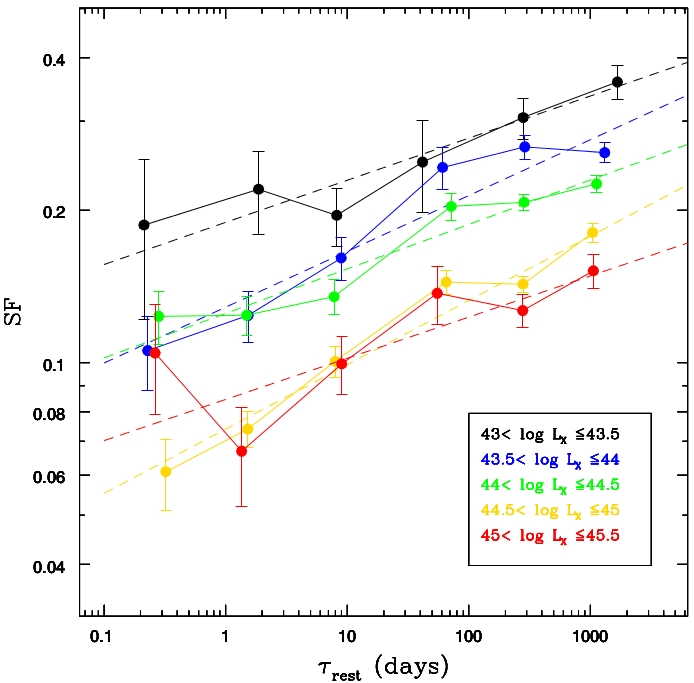}
\end{minipage}%
\begin{minipage}[b]{.5\linewidth}
\centering\includegraphics[width=8cm]{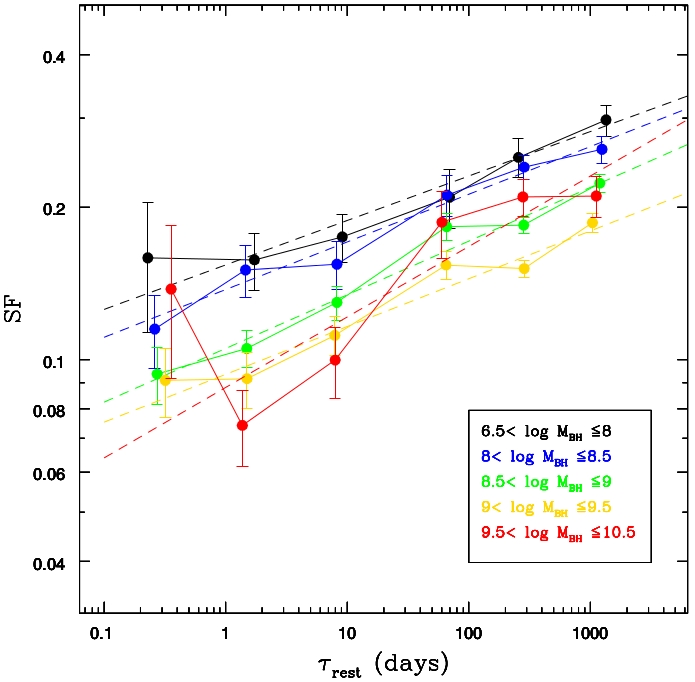}
\end{minipage}
\caption{The structure function in bins of X-ray luminosity $L_X$ and black hole mass $M_{BH}$ (data and continuous lines) with power-law fits (dashed lines).  {\it Left panel:} Black, $43<\log L_X\leq 43.5$; blue, $43.5<\log L_X\leq 44$; green, $44<\log L_X\leq 44.5$; yellow, $44.5<\log L_X\leq 45$; red, $45<\log L_X\leq 45.5$. {\it Right panel:} Black, $6.5\lesssim\log M_{BH}\leq 8$; blue, $8<\log M_{BH}\leq 8.5$; green, $8.5<\log M_{BH}\leq 9$; yellow, $9<\log M_{BH}\leq 9.5$; red, $9.5<\log M_{BH}\lesssim 10.5$.}
\end{figure}

\begin{figure}[h!]
\centering
\includegraphics[width=8cm]{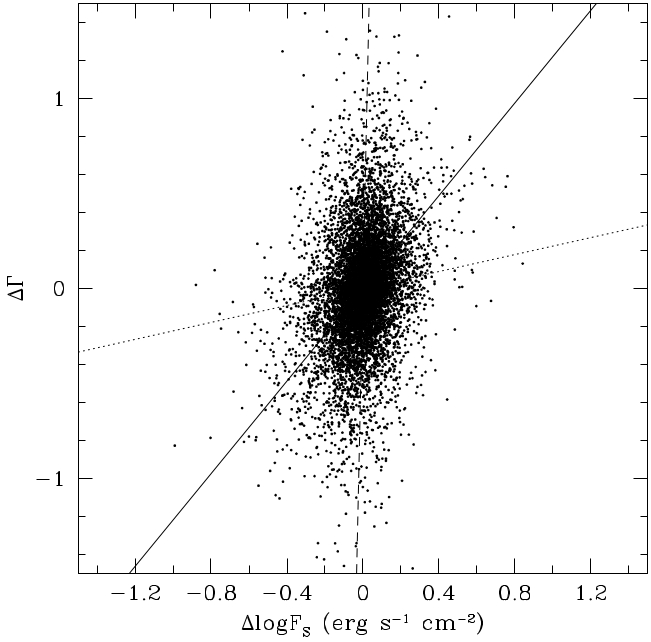}
\caption{The ensemble correlation between the variations of the photon index $\Delta\Gamma$ and the variations of the logarithmic flux $\Delta\log F$ are shown. The dotted line represents $\beta_{xy}$, the dashed line represents $\beta_{yx}$, while the bisector $\beta_{bis}$ is shown as a continuous line.}

\end{figure}

\begin{figure}[h!]
\begin{minipage}[b]{.5\linewidth}
\centering\includegraphics[width=8cm]{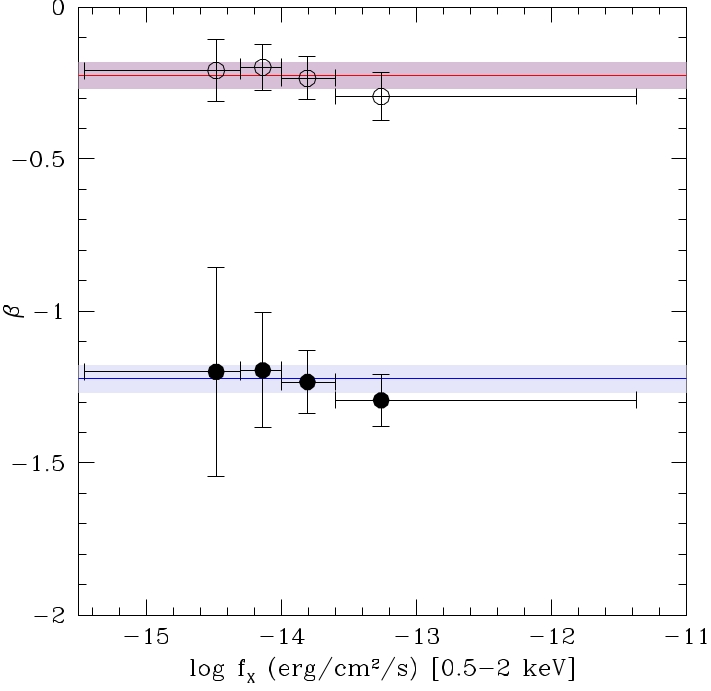}
\end{minipage}%
\begin{minipage}[b]{.5\linewidth}
\centering\includegraphics[width=8cm]{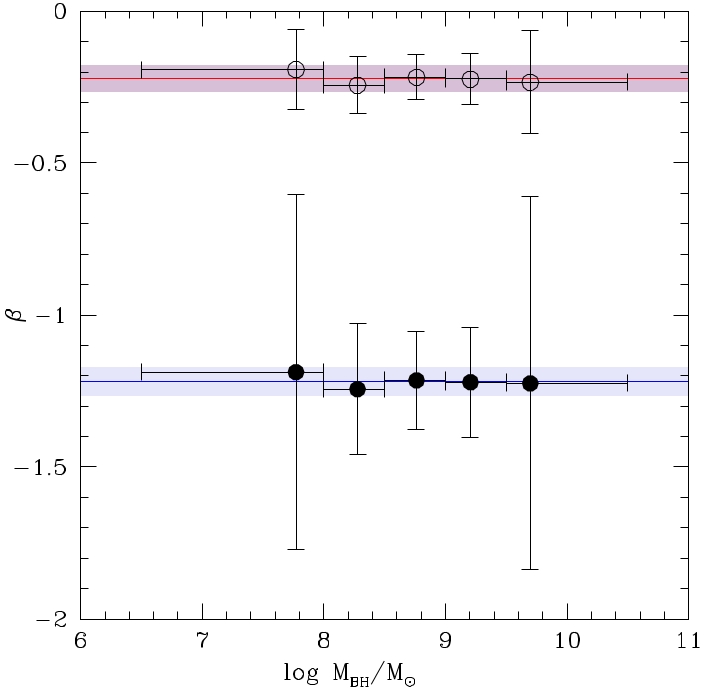}
\end{minipage}
\caption{The dependence of $\beta_{xy}$ and $\beta_{bis}$ in bins of soft X-ray flux {\it (left panel)} and black hole mass {\it (right panel)}. No dependence is found, and all the values of $\beta_{xy}$ (empty circles) and $\beta_{bis}$ (filled circles) are compatible with the corresponding ensemble values $\beta_{xy}=-0.22\pm0.04$ and $\beta_{bis}=-1.22\pm 0.05$, which are indicated respectively by the red and blue lines, and by the lighter bands for the uncertainties. The horizontal bars indicate bin widths, not errors.}
\end{figure}




\end{document}